\documentclass{PoS}

\usepackage[utf8]{inputenc} 
\usepackage[T1]{fontenc}  
\usepackage{amsmath,dsfont}

\usepackage{ifthen}
\def\eprinttmp@#1arXiv:#2 [#3]#4@{
\ifthenelse{\equal{#3}{x}}{\href{http://arxiv.org/abs/#1}{arXiv:#1}}{\href{http://arxiv.org/abs/#2}{arXiv:#2} [#3]}}
\newcommand{\eprint}[1]{\eprinttmp@#1arXiv: [x]@} 

\title{Quantum Simulation of Non-Abelian Lattice Gauge Theories}

\ShortTitle{Quantum Simulation of Non-Abelian Lattice Gauge Theories}

\author{\speaker{Michael Bögli}\\
{Albert Einstein Center for Fundamental Physics \\
Institute for Theoretical Physics, Bern University\\
Sidlerstr.\ 5, 3012 Bern, Switzerland.} \\
        E-mail: \email{boegli@itp.unibe.ch}}

\abstract{We use quantum link models to construct a quantum simulator for $U(N)$ and $SU(N)$ lattice gauge theories. These models replace Wilson's classical link variables by quantum link operators, reducing the link Hilbert space to a finite number of dimensions. We show how to embody these quantum link models with fermionic matter with ultracold alkaline-earth atoms using optical lattices. Unlike classical simulations, a quantum simulator does not suffer from sign problems and can thus address the corresponding dynamics in real time. Using exact diagonalization results we show that these systems share qualitative features with QCD, including chiral symmetry breaking and we study the expansion of a chirally restored region in space in real time.}

\FullConference{31st International Symposium on Lattice Field Theory LATTICE 2013\\
                 July 29 - August 3, 2013\\
                 Mainz, Germany}

\begin{document}

\section{Introduction}
Non-Abelian gauge theories are used to describe interactions in the standard model of particle physics. For example, the strong interaction gives rise to non-perturbative effects in the QCD $SU(3)$ gauge theory, including chiral symmetry breaking. 
Non-perturbative effects can in general not be derived analytically. Instead, one can use Monte Carlo based methods to obtain insights into these non-perturbative effects. However, sometimes Monte Carlo methods fail, for example, if we encounter a {sign problem}, which can have different origins. 
Non-equilibrium physics, for example, needs to be calculated in real time, which leads to a sign problem. Sign problems also appear in models of condensed matter systems, such as geometrically frustrated quantum antiferromagnets or various quantum spin liquids. 
To circumvent the sign problem, we can use quantum simulators. 
 
A quantum simulator is a system of isolated quantum objects, e.g.\ atoms. Since these systems embody the quantum nature already in their basic degrees of freedom, one does not encounter a sign problem. One way to build a quantum simulator is to use {optical lattices} \cite{bloch2012quantum}, where a set of laser beams is tuned to form  a periodic potential. Individual atoms then find the minima of this potential and thus arrange themselves in the lattice geometry. With other lasers one can encode or read out quantum information of the atoms, e.g.\ their nuclear spin. When constructing an optical lattice, one can make use of a rich experimental toolbox. Already several systems in condensed matter physics have been emulated in an optical lattice setup, including the bosonic Hubbard model \cite{greiner2002quantum}. Since one can perform measurements at arbitrary times, one can even address {dynamical questions}, with interesting potential applications in high energy or in condensed matter 
physics. 

It is most convenient to implement models with a finite-dimensional  Hilbert space in a quantum simulator. Quantum simulators for lattice models  with a local  $U(1)$  gauge symmetry with \cite{Kapit:2010qu,Banerjee.12.QLMU1,Zohar:2012ts} and without \cite{Buchler:2005fq,Zohar:2012ay,Tagliacozzo:2012vg} coupling to matter fields have already been constructed. Some of these constructions take advantage of the quantum link formulation
, which has a finite-dimensional Hilbert space per link. Quantum link models have been introduced for $U(1)$ and $SU(2)$ gauge groups \cite{Horn:1981kk,Orland:1989st,Chandrasekharan:1996ih} and extended to other gauge groups including the QCD gauge group $SU(3)$ \cite{Brower:1997ha,Brower:2003vy}. Based on \cite{Banerjee:2012xg}, here we propose a construction of a quantum simulator for non-Abelian quantum link models with a $U(N)$ gauge symmetry coupled to staggered fermions. Other implementations of non-Abelian gauge theories in optical lattices have been described in \cite{Tagliacozzo:2012df,Zohar:2012xf}. We will see, already a simple realization in ($1+1$)~dimensions shares non-trivial features of QCD, including confinement or spontaneous chiral symmetry breaking%
. 
This construction can be realized with ultracold alkaline-earth atoms, such as $^{87}$Sr or $^{193}$Yb, in an optical lattice. 

\section{Quantum Link Models}
\label{sec:qlm}
\subsection{From Wilson's Lattice Gauge Theory to Quantum Link Models}
\begin{figure}[thb]
 \centering
 \setlength{\unitlength}{0.5cm}
\begin{picture}(12,2)(-2,-1)
\put(0,0){\circle*{0.2}}
\put(8,0){\circle*{0.2}}
\put(-1,0){\line(1,0){10}}
\put(-0.2,-0.7){$x$}
\put(7.8,-0.7){$y = x$}
\put(3.4,-0.9){$U_{x,y}^{ij}$}
\put(3.4,0.4){$E_{x,y}$}
\put(1.0,0.4){$L_{x,y}^a$}
\put(5.8,0.4){$R_{x,y}^a$}
 \end{picture} 
\caption{The quantum link operator $U_{x,y}$ and the electric flux operators defined on the link $x, y$. }
\label{fig:link}
\end{figure}
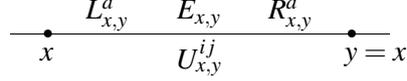

When formulating Wilson's $SU(N)$ or $U(N)$  lattice gauge theory in a Hamiltonian formalism, we work with the canonical conjugate operators of the entries of the gauge link matrices $U^{ij}_{x,y}$ and $U^{\dagger ij}_{x,y}$. These operators are the $SU(N)$ electric and magnetic flux operators  $L_{x,y}^a$ and $R_{x,y}^a$ ($a=1,\ldots, N^2-1)$, which are associated with the left and right end of the link (see figure \ref{fig:link}). The two operators generate an $SU(N)_L \otimes SU(N)_R$ algebra on each link (see equation (\ref{eq:commute0})). There is also an operator $E_{x,y}$, representing the Abelian $U(1)$ electric flux. These operators obey
\begin{align}
[R^a, R^b] &= 2if^{abc}R^c,  & [ R^a, U^{ij}] &=  U^{ik}\lambda^a_{kj}, & [ L^a, U^{ij}] &= -\lambda^a_{ik} U^{kj}, & [E, U^{ij}]   &= U^{ij}, \nonumber \\ 
[L^a, L^b] &= 2i f^{abc} L^c, &[R^a, L^b] &= 0,  &  [E, R^a] &= 0,  &[E, L^a] &= 0, \label{eq:commute0} 
\end{align}
while operators associated with different links commute. Here, $\lambda^a$ are the $SU(N)$ Gell-Mann matrices and $f^{abc}$ are the $SU(N)$ structure constants.

Quantum link models are an extension of ordinary Wilson type lattice gauge theory.
The idea is to quantize the elements of the link matrices in a similar fashion as we quantize a classical spin by forming a quantum spin operator 
acting in a Hilbert space. In a similar way, we can introduce quantum operators for the elements of the gauge link matrix  $U_{x,y}^{ij}$ by associating each element with a non-commuting operator, which is called a quantum link operator. We thus give up the commutativity of the entries of the gauge link matrices $U_{x,y}^{ij}$ and $U_{x,y}^{\dagger ij}$
\begin{align}
[U^{ij}, U^{\dagger kl}] &= 2(\delta_{ik}  \lambda_{jl}^{*a} R^a - \delta_{jl}\lambda_{ik}^a L^a + 2\delta_{ik}\delta_{jl} E),
&[U^{\dagger ij}, U^{\dagger kl}] &= [U^{ij}, U^{kl}]  = 0 \label{eq:commute2}.
\end{align}

Equation (\ref{eq:commute0}) and (\ref{eq:commute2}) define an embedding $SU(2N)$ algebra, where $L_{x,y}^a$ $R_{x,y}^a$, $E_{x,y}$ and combinations of $U^{ij}_{x,y}$ and $U^{\dagger ij}_{x,y}$ are the generators. By choosing an irreducible representation of $SU(2N)$ on each link, the link Hilbert space is only finite-dimensional. 

\subsection{Rishon Representation}
For simplicity, we only discuss a $(1+1)$-dimensional system. Since we want to implement the quantum links on an optical lattice, it is convenient to introduce the so called rishon representation. Rishons are fermions associated with the left and right end of a link (see figure \ref{fig:rishons}). The creation $c_{x,\pm }^{\dagger i}$ and annihilation operators  $c_{x,\pm }^i$ obey the usual anti-commuting relations
\begin{equation}
 \{ c_{x,\pm }^i, c_{y,\pm }^j \} =  \{ c_{x,\pm }^{\dagger i}, c_{y,\pm }^{\dagger j}\} = 0, \qquad \{ c_{x,\pm}^i, c_{y,\pm}^{\dagger j} \}= \delta_{xy}\delta_{\pm, \pm} \delta_{ij}, 
\end{equation}
where $i,j$ are color indices $i,j = 1, \ldots, N$. 
It is possible to express all generators of the $SU(2N)$ algebra in terms of these rishons
\begin{align}
 R^a_{x,x+1} &= c_{x+ 1, - }^{\dagger i}\ \lambda_{ij}^a\ c_{x+ 1, -}^j, 
 & L^a_{x,x+1} &= c_{x, +}^{\dagger i}\ \lambda_{ij}^a\ c_{x, +}^j, \nonumber\\
 E_{x,x+1} &= \frac 12 \left(c_{x+ 1, - }^{\dagger i}c_{x+ 1, - }^i - c_{x, +}^{\dagger i}\ c_{x, +}^i \right), 
 & U_{x,x+1}^{ij} & = c_{x, +}^{ i}\ c_{x+ 1, - }^{\dagger j}. 
\end{align}
One can check that the commutation relations  (\ref{eq:commute0}) and (\ref{eq:commute2}) are indeed satisfied. 

\subsection{The Hamiltonian}
We introduce staggered fermions with creation $\psi^{\dagger i}_x$ and annihilation operators $\psi_x^i$ obeying the usual anti-commuting relations
\begin{equation}
 \{ \psi_{x }^i, \psi_{y}^j \} =  \{ \psi_{x}^{\dagger i}, \psi_{y }^{\dagger j}\} = 0, \qquad \{ \psi_{x }^i, \psi_{y}^{\dagger j} \}= \delta_{xy} \delta_{ij}.
\end{equation}
The Hamiltonian couples the quantum links to fermions of mass $m$
\begin{align}
 H &= -t \sum_{x } \left( \psi_x^{i\dagger} U^{ij}_{x, x+1}\psi_{x+1}^j + \text{h.c.} \right) + m\sum_x (-1)^x \psi_x^{i\dagger}\psi_{x}^i \nonumber \\
 &= -t \sum_{x } \left( \psi_x^{i\dagger}  c_{x,+}^i \ c_{x+1,-}^{j\dagger}\psi_{x+1}^j + \text{h.c.} \right) + m\sum_x (-1)^x \psi_x^{i\dagger}\psi_{x}.
\end{align}

It turns out that the rishon representation is very useful. We see that a hop of a fermion from $x+1$ to $x$ simultaneously induces a hop of a rishon from $x$ to $x+1$. Therefore this Hamiltonian describes the hopping of different kinds of fermions, and we have a Hubbard-like model with some more constraints: In particular, the total number of rishons per link is conserved. This is also the case in higher-dimensions when the Hamiltonian also contains a plaquette term.

\subsection{Symmetries} 
As already discussed, we consider an $SU(N)\times U(1) = U(N)$ gauge symmetry. The generators of the $SU(N)$ gauge transformation $G^a_x$ ($a = 1, \ldots, N^2-1$) obey $[G^a_x, G^b_y] = 2i \delta_{xy}f^{abc}G^c_x$. Together with the generator of the additional $U(1)$ symmetry $G_x$  they can be expressed in terms of the flux operators $L^a$, $R^a, E$, and the fermion fields $\psi^i$, $\psi^{\dagger i}$
\begin{align}
  SU(N): & &G_x^a &= \psi_x^{\dagger i} \lambda_{ij}^a \psi_x^j + \left(L_{x,x+1}^a+  R_{x-1,x}^a \right), \nonumber \\
   U(1): & &G_x &= \psi_x^{\dagger i} \psi_x^i + (E_{x-1, x} - E_{x, x + 1}).
\end{align}
We see that these definitions are correct by using the commutation relation derived from (\ref{eq:commute0})
\begin{equation}
 [U_{x,x+1}^{ij}, G_y^a] = \delta_{xy} \lambda^a_{ik} U_{x,x+1}^{kj} -  \delta_{x+1, y} U_{x,x+1}^{ik}\lambda^a_{kj}. \label{eq:commute1}
\end{equation}
With this we can check that the quantum link operators $U^{ij}_{x,y}$ and the fermion fields $\psi^i_x$ transform properly under gauge transformations 
\begin{align}
SU(N): & & V^\dagger U_{x,y}^{ij} V &= \bigl[ \exp \bigl(i \alpha_x^a \lambda^a \bigr) \bigr]^{ik} U_{x,y}^{kl} \bigl[ \exp \bigl(-i \alpha_{y}^a \lambda^a \bigr) \bigr]^{lj}, &V^\dagger \psi_x^i V & =  \left[\exp \bigl(i \alpha_x^a \lambda^a \bigr)\right]^{ij} \psi_x^j, \nonumber \\
 U(1): & &W^\dagger U_{x,y}^{ij} W &= \exp \bigl(i \alpha_x\bigr)  U_{x,y}^{ij} \exp \bigl(-i \alpha_{y}\bigr) , &W^\dagger \psi_x^i W& =  \exp \bigl(i \alpha_x \bigr) \psi_x^i. \label{eq:gaute_transf}
\end{align}
Here,   $V=\prod_x \exp \left(i \alpha_x^a G_x^a \right)$ and $W=\prod_x \exp \left(i \alpha_x^a G_x \right)$ are unitary transformations implementing general gauge transformations. Note the difference between the two objects $G^a_x$  and $\lambda^a$, which share the same commutation relation: while $\lambda^a$ acts as a matrix on the color indices, $G^a_x$ generates unitary transformation in the entire Hilbert space. These transformations are the correct $SU(N)$ or $U(N)$ gauge transformations and leave the Hamiltonian invariant $[G^a_x, H] = [G_x, H] = 0$. 
The additional $U(1)$ gauge symmetry can later be explicitly broken in order to reduce the $U(N)$ gauge symmetry to $SU(N)$ (see e.g.\ supplementary material of \cite{Banerjee:2012xg}). 



Furthermore, the Hamiltonian has certain global symmetries: spatial translations, charge conjugation, parity, baryon number symmetry (in the $SU(N)$ case), as well as a $\mathbb Z_2$ chiral symmetry. All these symmetries are explained in detail in \cite{Banerjee:2012xg}. Since we want to focus on chiral symmetry breaking, here we only discuss the chiral symmetry. A chiral symmetry transformation $\chi$ for a 1-dimensional system corresponds to a shift by one lattice spacing. Therefore the transformation rules are \vspace{-4mm}
\begin{align}
  \chi: & &^\chi U_{x,x+1} &= U_{x+ 1,x+2} , &^\chi \psi_x & =  \psi_{x + 1}.
\end{align}
As we can see, the mass term in the Hamiltonian breaks the chiral symmetry explicitly. Therefore we have an exact chiral symmetry only in the case of massless fermions, $m=0$. 

\subsection{Implementation in an Optical Lattice}
\begin{figure}[htb]
 \centering
 \vspace{-4mm}
 \includegraphics[width=0.35\textwidth]{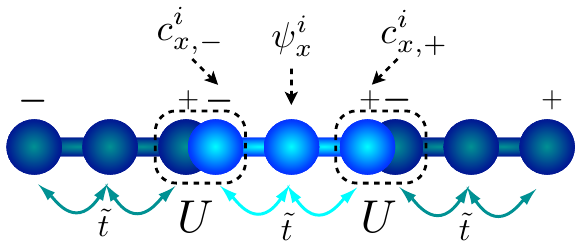}
 \caption{A superlattice with rishon and fermion sites. Quantum link operators $U$ can be represented in terms of rishon operators $c$, $c^\dagger$.}
 \label{fig:rishons}
\end{figure}
The rishon representation allowed us to rewrite this model as a system of hopping fermions. The implementation in a quantum simulator can be realized with a single species of alkaline-earth atoms \cite{Banerjee:2012xg} representing either the rishons or the staggered fermions depending on their location in an optical superlattice. Instead of an actual hop of a fermion from $x+1$ to $x$ and a hop of the rishon from $(x,+)$ to $(x+1,-)$, it is more convenient to move the alkaline-earth atom from $x+1$ to the rishon position $(x+1, -)$, while moving another atom from the  rishon position $(x, +)$ to the fermion position $x$. An optical superlattice guarantees this behavior by restricting some atoms only to move within the bright part in figure \ref{fig:rishons} and some atoms only within the darker region. 

To implement this system in an optical lattice setup one needs only one species of fermions, which hop between neighboring sites, keeping the total number of rishons per link constant. In our proposed implementation we use alkaline-earth atoms (e.g.\ $^{87}$Sr or $^{173}$Yb).  The color degrees of freedom are encoded in the Zeeman levels of these atoms, where due to their nuclear spin $I$, we have an $SU(2I+1)$ symmetry in which the gauge group $SU(N)$ or $U(N)$ can be embedded. 

\section{Exact Diagonalization Results}

\subsection{Explicit Definition of the Hilbert Space}
To investigate whether this model actually shows interesting physics, we now present some exact diagonalization results for a $(1+1)$-dimensional system with a $U(2)$ gauge symmetry. In order to do this, we introduce the gauge invariant operators $M_x = \psi_x^{\dagger i} \psi_x^i$ and $Q_{x, \pm } = c_{x, \pm }^{\dagger i}\psi_x^i$. Using these we rewrite the $(1+1)$-d Hamiltonian to
\begin{align}
 H &= -t \sum_{x } \left( \psi_x^{i\dagger}  c_{x,+}^i \ c_{x+1,-}^{j\dagger}\psi_{x+1}^j + \text{h.c.} \right) + m\sum_x (-1)^x \psi_x^{i\dagger}\psi_{x} \nonumber \\
 &= -t \sum_{x } \left( Q_{x, +}^\dagger  Q_{x+1,-}+ \text{h.c.} \right) + m\sum_x (-1)^x M_x.
 \end{align}
If we work in a representation with only one rishon per link, we see that only four states at each site $x$ are gauge invariant (i.e.\ $G^a_x |\psi\rangle_x = 0$)
\begin{align}
 | 1 \rangle_x & = \frac 1{\sqrt 2} \left( c_{x, -}^{\dagger 1} c_{x, +}^{\dagger 2} - c_{x, -}^{\dagger 2} c_{x, +}^{\dagger 1} \right) |0\rangle, & | 2 \rangle_x & = \frac 1{\sqrt 2} \left( c_{x, -}^{\dagger 2} \psi_{x}^{\dagger 1} - c_{x, -}^{\dagger 1} \psi_{x}^{\dagger 2} \right) |0\rangle, \nonumber \\
 | 3 \rangle_x & = \frac 1{\sqrt 2} \left( c_{x, +}^{\dagger 2} \psi_{x}^{\dagger 1} - c_{x, +}^{\dagger 1} \psi_{x}^{\dagger 2} \right) |0\rangle,  & | 4 \rangle_x & =  \psi_{x}^{\dagger 2} \psi_{x}^{\dagger 1}|0\rangle.
\end{align}
Now we write the operators $Q_{x, \pm }, M_x$ in the basis of the gauge invariant states $\{ |1\rangle, |2\rangle, |3\rangle, |4\rangle \}$
\begin{equation}
 M_x = \left[\begin{array}{cccc}
  0 & 0 & 0 & 0 \\
  0 & 1 & 0 & 0 \\
  0 & 0 & 1 & 0 \\
  0 & 0 & 0 & 2 
     \end{array} \right], \qquad Q_{x,+} = \left[\begin{array}{cccc}
  0 & 1 & 0 & 0 \\
  0 & 0 & 0 & 0 \\
  0 & 0 & 0 & \sqrt 2  \\
  0 & 0 & 0 & 0 
     \end{array} \right], \qquad Q_{x,-} = \left[\begin{array}{cccc}
  0 & 0 & 1 & 0 \\
  0 & 0 & 0 & \sqrt 2 \\
  0 & 0 & 0 & 0  \\
  0 & 0 & 0 & 0 
     \end{array} \right].
\end{equation}
Because we have four states per link, we see that the dimension of the Hilbert space scales as $4^L$ with the number $L$ of lattice points. However, there is a reduction of the Hilbert space: since the rishon is either on the left or on the right end of the link, one is restricted to two states for each neighbor. Therefore the dimension of the total Hilbert space only scales as $2^L$. 

\subsection{Exact Diagonalization Results}
\begin{figure}[htb]
 \centering
 \vspace{-4mm}
 \includegraphics[trim=0cm 2mm 0cm 4mm, clip=true,width=0.7\textwidth]{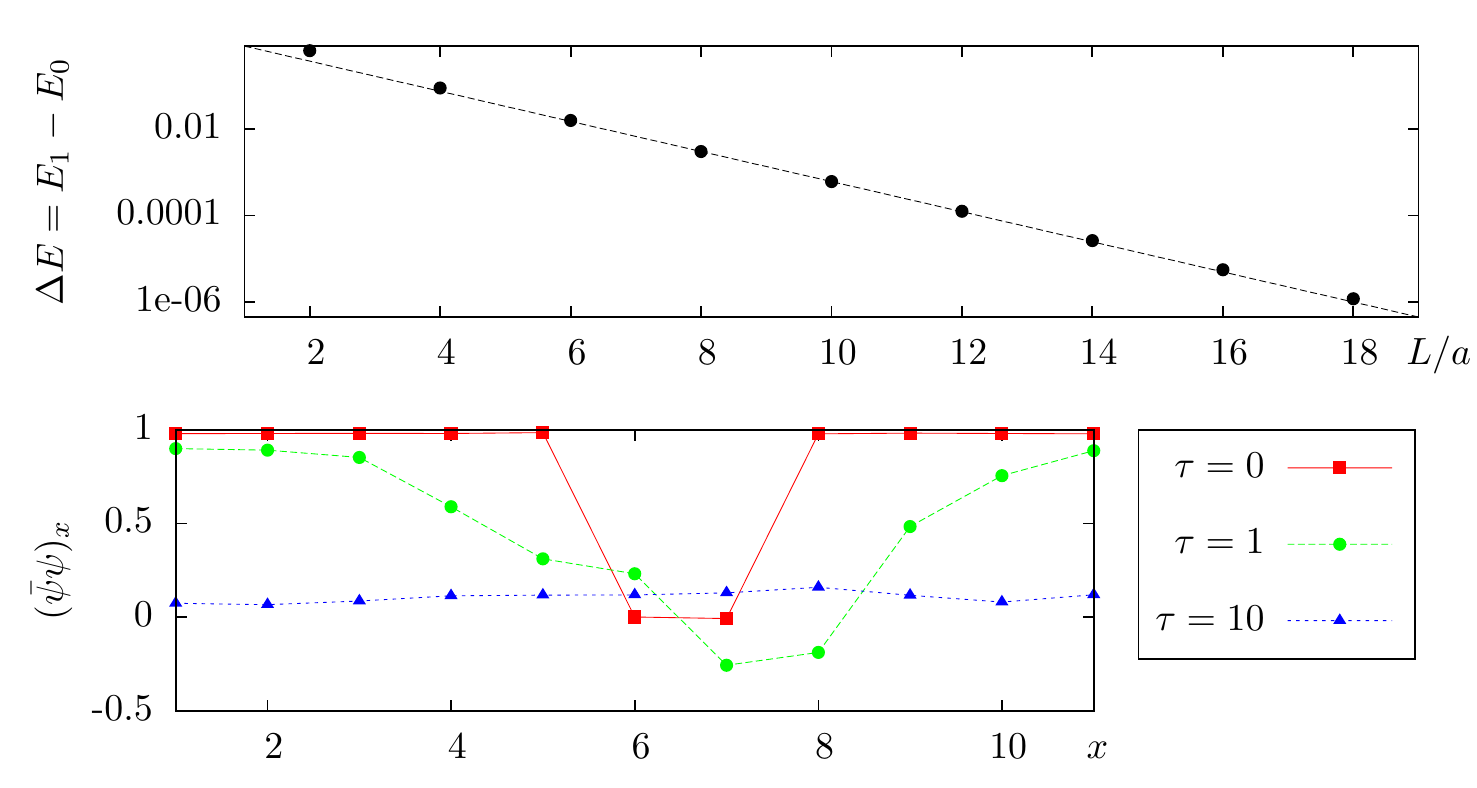}
 \caption{Chiral symmetry breaking: (top) Scaling of the energy difference of almost degenerate vacua, (bottom) spatial dependance of the chiral order parameter for various real times $\tau = 0, 1, 10$. }
 \label{fig:ed_result}
\end{figure}
A system, in which a discrete $\mathbb Z_2$ chiral symmetry is broken spontaneously, will show an ``almost degenerate'' spectrum in a finite volume. This means that the spectrum contains always pairs of states, which have almost the same energy. These pairs of states are related by the spontaneously broken chiral symmetry  and the energy difference $\Delta E$ between those two states decreases exponentially with the volume $L$. In figure \ref{fig:ed_result} (top) we show these energy differences, which indeed confirms the chiral symmetry breaking. 

The expansion of a chirally restored hot-spot is also calculated in real time. To do this we start in an initial configuration, where the chiral symmetry is broken everywhere (background vacuum) except at two points (hot-spot) where chiral symmetry is restored. The hot-spot mimics, for example, the quark-gluon plasma in a heavy ion collision. After certain time-intervals ($\tau = 0, \tau = 1, \tau = 10$) we measure the order parameter of the chiral symmetry breaking
\begin{equation}
 (\bar \psi\psi)_x = (-1)^x \cdot ( 1- M_x).
\end{equation}
In the plot shown in figure \ref{fig:ed_result} (bottom), we see how with increasing time the symmetric phase is spreading out on the lattice.

\section{Conclusion and Outlook}
We proposed a construction for a quantum simulator of a $U(N)$ lattice gauge theory in an optical lattice setup. This is realized by using quantum links coupled to staggered fermions. The rishon representation allowed us to rewrite the Hamiltonian in a Hubbard-like way, with only fermions hopping on the lattice. This is then used to implement this model using ultra-cold alkaline-earth atoms in an optical lattice. We showed exact diagonalization results of a $(1+1)$-d $U(2)$ quantum link model. Already in a simple model like this, we were able to demonstrate interesting physics like the real-time dynamics of chiral symmetry restoration across the phase transition. 

A next step will be to simulate models with finite baryon density. Later it would also be interesting to simulate phenomena like baryon superfluidity, color superconductivity at high densities and ``nuclear collisions''. A long-term goal is to quantum simulate full QCD in real time.

\subsection*{Acknowledgments}
I thank all my collaborators from Innsbruck, Ulm, and Bern, D. Banerjee, M. Dalmonte, E.~Rico, P. Stebler, U.-J. Wiese, and P. Zoller for the interesting work we have done together. This work is supported by the Schweizerischer Nationalfonds. 

\bibliographystyle{boegli-no_titles}
\bibliography{quantum_simulation_LGT}

\end{document}